\documentclass[aps,showpacs,preprintnumbers,nofootinbib,amsmath,amssymb]{revtex4-1}
\usepackage{slashed,color,amsmath,amssymb,mathrsfs}
\usepackage{graphicx}
\usepackage{ulem}
\usepackage{hyperref}
\usepackage{longtable}
\usepackage{array}

\allowdisplaybreaks

\begin{document}
\preprint{}
\title{Exothermic isospin-violating dark matter after SuperCDMS and CDEX}

\author{Nan Chen$^1$}
\author{Qing Wang$^{1,2,3}$}
\email[Corresponding author.]{wangq@mail.tsinghua.edu.cn}
\affiliation{$^1$Department of Physics, Tsinghua University, Beijing 100084, P. R. China\\
$^2$Center for High Energy Physics, Tsinghua University, Beijing 100084, P. R. China\\
$^3$Collaborative Innovation Center of Quantum Matter, Beijing 100084, P. R. China}
\author{Wei Zhao$^4$}\author{Shin-Ted Lin $^5$}\author{Qian Yue$^4$}\author{Jin Li$^4$}
\affiliation{$^4$Key Laboratory of Particle and Radiation Imaging (Ministry of Education) and Department of Engineering Physics,
Tsinghua University, Beijing 100084, P. R. China\\
$^5$College of Physical Science and Technology, Sichuan University, Chengdu 610064, P. R. China}

\begin{abstract} We show that exothermic isospin-violating dark matter (IVDM) can make the results of the latest CDMS-Si experiment consistent with recent null experiments, such as XENON10, XENON100, LUX, CDEX, and SuperCDMS, whereas for the CoGeNT experiment, a strong tension still persists. For CDMS-Si, separate exothermic dark matter or isospin-violating dark matter cannot fully ameliorate the tensions among these experiments; the tension disappears only if exothermic scattering is combined with an isospin-violating effect of $f_n/f_p=-0.7$. For such exothermic IVDM to exist, at least a new vector gauge boson (dark photon or dark Z') that connects SM quarks to Majorana-type DM particles is required.
 \end{abstract}
 \maketitle
Low-mass dark matter (DM) in the GeV energy region is currently the main topic of DM searches. On the one hand, some direct-detection experiments have claimed to have observed low-energy recoil events in excess of known backgrounds; these include DAMA \cite{DAMA1,DAMA2,DAMA3,DAMA4}, CoGeNT \cite{Cogent1,Cogent2,Cogent3,Cogent4}, and CRESST-II \cite{CRESSTII}\footnote{A possible excess over the background reported for the previous run (from 2009 to 2011) has not been confirmed in the upgraded CRESST-II detector, with an exposure of 29.35~kg live days collected in 2013\cite{CRESSTIInew}.}, and the latest such result is the positive CDMS-Si signal \cite{CDMSSi}. These excesses, if interpreted in terms of DM particles elastically scattering off target nuclei, may imply the existence of light DM particles with a mass of $<10$~GeV and a scattering cross section of approximately $10^{-41} \sim 10^{-40}\mathrm{cm}^2$. However, many other experiments, such as CDMS-II \cite{CDMSII1,CDMSII2,CDMSII3,CDMSlite}, XENON10/100 \cite{XENON10,XENON1001,XENON1002,XENON1003}, SIMPLE \cite{SIMPLE}, TEXONO \cite{TEXONO1,TEXONO2}, CDEX \cite{CDEX1}, LUX \cite{LUX}, the latest SuperCDMS \cite{SuperCDMS}, and CDEX \cite{CDEX2}, have reported null results in the same DM mass range. The serious conflict between these two completely different sets of results contrasts sharply with the situation in particle physics collider experiments, where all data appear to be in harmony with Standard Model (SM) predictions and, to date, no evidence of new physics beyond the SM has been observed. These tensions in the direct detection of DM are a strong motivation driving further investigations, which seek a deeper understanding either of the direct-detection experiments or of present theoretical interpretations of the experimental results. We may also treat the reconciliation of these contradictory experimental results as a guide for the identification of certain properties of the DM particle.

From the theoretical viewpoint, we need to establish whether there exists some mechanism to reconcile the present tension. If so, then we are closer to the discovery of the DM particle in particle physics experiments; if not, then experimentalists must conduct more complex background analysis to extract additional events from the observed signals. Note that most of the experimentally detected signals have been recorded using target materials that are different from those used in the experiments in which the null results have been obtained; the exception is CoGeNT, for which the target material is Ge, which is also used in CDMS-II, TEXONO, CDEX, and SuperCDMS. Although a contingent of CoGeNT researchers has recently released an improved analysis of 3.4 years of CoGeNT data \cite{Cogent5}, exhibiting a close similarity to previously reported results \cite{Cogent3,Cogent4} on the annual modulation, though different and weaker, questions remain regarding their surface event analysis \cite{QuestionCogent}. By contrast, J.H. Davis has announced that the DAMA result can be fitted using neutrons from muons and neutrinos instead of DM \cite{DAMAalternativeExplanation}. Shortly following this announcement, comments appeared claiming that the effect from muons and neutrinos is negligible\cite{DAMAcomment1,DAMAcomment2}. R. Foot has attempted to use MeV-order DM scattering off electrons to explain the DAMA result\cite{DAMAalternativeExplanation1}. Recently, some researchers have used MeV-range axion-like particles to explain the DAMA signal\cite{DAMAalternativeExplanation2}; soon after, others claimed that this
model has already been ruled out by many orders of magnitude based on existing experimental results\cite{DAMAcomment3}. Considering that CRESST-II's signal has already disappeared \cite{CRESSTIInew}, if we ignore these debatable CoGeNT and DAMA results, then, in our analysis, there are two popular theoretical scenarios that are able to ameliorate the tensions between the remaining CDMS-Si results and the other null experiments in regard to the details of the different structures of their target nuclei. One is isospin-violating DM (IVDM)\cite{IVDM1,IVDM2,IVDM3,IVDM4,IVDM5,IVDM6,IVDM7,IVDM8}, wherein the DM particle couples to protons and neutrons with different strengths; possible destructive interference resulting from these two couplings can weaken the bounds of XENON10/100 and move the signal regions of DAMA and CoGeNT closer to each other \cite{IVDM5,IVDM6}. To reconcile the data from DAMA, CoGeNT, and XENON10/100, a large destructive interference is required; this interference is dependent on the ratio of the spin-independent scatterings of the couplings of the DM particle to the neutron ($f_n$) and to the proton ($f_p$), which must be of order $f_n/f_p\approx-0.7$ \cite{IVDM5}. However, from indirect DM searches, such as the antiproton flux measured by BESS-Polar II, the relevant couplings for IVDM have been found to be severely constrained \cite{IVDM9,IVDM10}. Furthermore, after the appearance of the LUX data \cite{LUX}, it was observed \cite{tension1,tension2,tension3} that LUX and CDMS-Si are now in tension even for IVDM. The other possible scenario is to go beyond conventional elastic scattering and consider whether DM scatters inelastically to a lower mass state; such DM is termed exothermic \cite{ExcitingDM1,ExcitingDM2,ExcitingDM3,ExcitingDM4, ExcitingDM5,ExcitingDM6,ExcitingDM7}. For a sufficiently long-lived heavier state, there must be sufficient numbers of such DM particles in the vicinity of the Earth to produce a signal in direct-detection experiments, and the splitting cannot be too large. We consider $\delta\leq 200$ keV for appropriately small splittings; in that case, the only available decays are to neutrinos or photons. If couplings in the SM occur through the kinetic mixing of a dark-sector gauge boson with the SM gauge bosons, then the lifetimes are longer than the age of the universe \cite{ExcitingDM2,Lifetime1}. By choosing a mass splitting between the DM excited and de-excited states of approximately $\delta\sim-200$ keV, \cite{ExcitingDM8,ExcitingDM9} one can accommodate both the LUX and CDMS-Si results and simultaneously account for the high- and low-energy events.

Although the exothermic DM model succeeds in relaxing the tension between LUX and CDMS-Si \cite{ExcitingDM8,ExcitingDM9}, this model has not been considered in the aftermath of the latest results from SuperCDMS \cite{SuperCDMS} and CDEX \cite{CDEX2}. Such an analysis is the topic of the present paper. Our objective in the study is to examine the consistency of the SuperCDMS and CDEX null results with the excess CDMS-Si result by implementing the two scenarios mentioned above. The CoGeNT and DAMA results will also be considered as references in our discussion, although the interpretations thereof are still a subject of debate. Because exothermic DM, unlike endothermic DM, for which the DM scatters inelastically to a higher mass state, can reduce the relative modulation amplitude, the tension with the CoGeNT result is not expected to be reduced, and we shall see later that the DAMA result even shrinks to zero. Our strategy is first to apply exothermic DM to the SuperCDMS and CDEX results to confirm whether these more stringent experiments are consistent with the CDMS-Si result. If so, then exothermic DM becomes a unique type of DM that is consistent with all existing (except CoGeNT and DAMA) direct-detection experiments; if not, we will add in the IVDM effect and evaluate the results. As mentioned above, the IVDM model is already in tension with the results of LUX and CDMS-Si, and therefore, relying solely on IVDM to ameliorate the tensions among the different experiments is impossible; however, we can combine this mechanism with the exothermic DM model to further reduce these tensions.

For SuperCDMS, its latest result has recently been reported \cite{SuperCDMS}, in which the data obtained during 577 kg-days of exposure were analyzed for WIMPs of mass $<30~\mathrm{GeV}/c^2$ with a blinded signal region. Eleven events were observed once the analysis was complete. The authors set an upper limit on the spin-independent WIMP-nucleon cross section of $1.2\times10^{-42}\mathrm{cm}^2$ at $8~\mathrm{GeV}/c^2$. In the meantime, CDEX already published its latest null results \cite{CDEX2} for 53.9 kg-days of data.

To interpret the above experimental results in terms of inelastic scattering, we note that exothermic DM particles are those DM particles $\chi_1$ of mass $m_1$ that inelastically down-scatter to DM particles $\chi_2$ of mass $m_2$ from a nucleus $N$ as follows: $\chi_1+N\rightarrow \chi_2 +N$. The requisite velocity to produce a nuclear recoil of energy $E_R$ is
\begin{eqnarray}
v_{\mathrm{min}}=\frac{1}{\sqrt{2E_Rm_N}}|\delta+\frac{m_NE_R}{\mu}|,\hspace*{2cm}\delta\equiv m_2-m_1<0\label{vmin},
\end{eqnarray}
where $\mu$ is the reduced mass of the DM-nucleon system. Up-scattering ($\delta>0$) is more prevalent from heavy nuclei, whereas down-scattering ($\delta<0$) is more prevalent from light nuclei, where the energy of the recoiling nucleus is peaked near a scale that is proportional to the splitting between the dark matter states and is inversely proportional to the nuclear mass. Consequently, nuclear recoils caused by exothermic DM ($\delta<0$) are more visible in experiments with light nuclei and low thresholds. Given the lightness of Si with respect to Xe and Ge, down-scattering is one avenue for explaining the CDMS-Si data while remaining consistent with the null XENON, LUX, SuperCDMS, and CDEX searches.
Figure 1 shows a plot of the elastic-scattering (corresponding to $\delta=0$) results from the CoGeNT, DAMA and CDMS-Si signal regions alongside the null results of XENON100, XENON10, LUX, SuperCDMS, and CDEX. For the DAMA experiment, it has been noted \cite{DAMA6} that nuclei recoiling along the characteristic axes or planes of
the crystal structure may travel large distances without colliding with other nuclei. This means that recoils that undergo such ion channeling have quenching factors of $Q_T\approx1$. We consider the cases both with and without this ion-channeling effect. The null experiments of XENON100, LUX and SuperCDMS are in strong tension with the CoGeNT, DAMA (both with and without the ion-channeling effect), and CDMS-Si results.
\begin{figure} \begin{center}
\scalebox{0.4}{\includegraphics{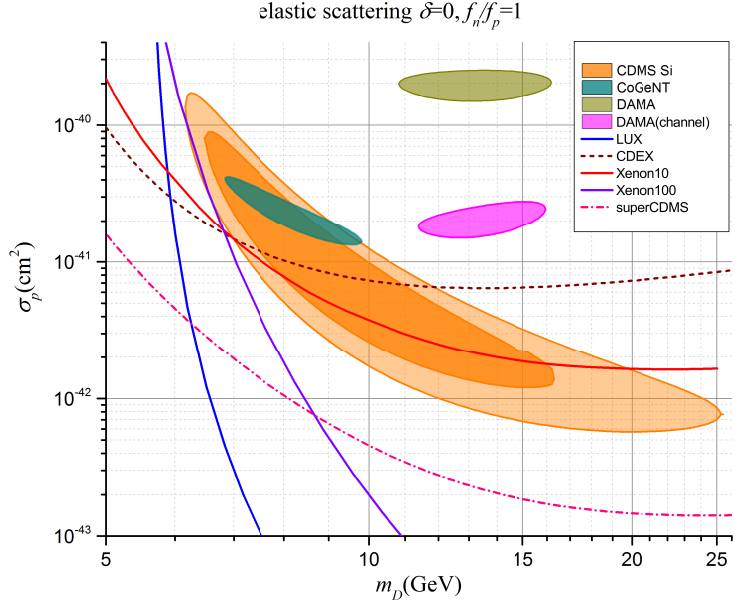}} \end{center} \vspace*{-0.5cm}
\caption{Elastic-scattering results without isospin violation. The exclusion lines for LUX (solid blue), CDEX (dashed brown), XENON10 (solid red), XENON100 (solid purple), SuperCDMS (dash-dotted magenta) are all at the 90$\%$ CL and are superimposed over the 68$\%$ (dark yellow) and 90$\%$ (light yellow) CL CDMS-Si best-fit regions, the 95$\%$ (dark cyan) CL CoGeNT best-fit region and the 95$\%$ CL DAMA best-fit regions without ion channeling (dark yellow) and with ion channeling (magenta).} \end{figure}

In Fig.~2, we consider the results for the inelastic scattering of exothermic DM (corresponding to $\delta<0$) from the CoGeNT and CDMS-Si signal regions along with the null results from XENON100, XENON10, LUX, SuperCDMS, and CDEX. The left panel corresponds to $\delta=-50$ keV, and the right panel corresponds to $\delta=-200$ keV.
\begin{figure} \begin{flushleft}
\hspace*{-1.3cm}\scalebox{0.4}{\includegraphics{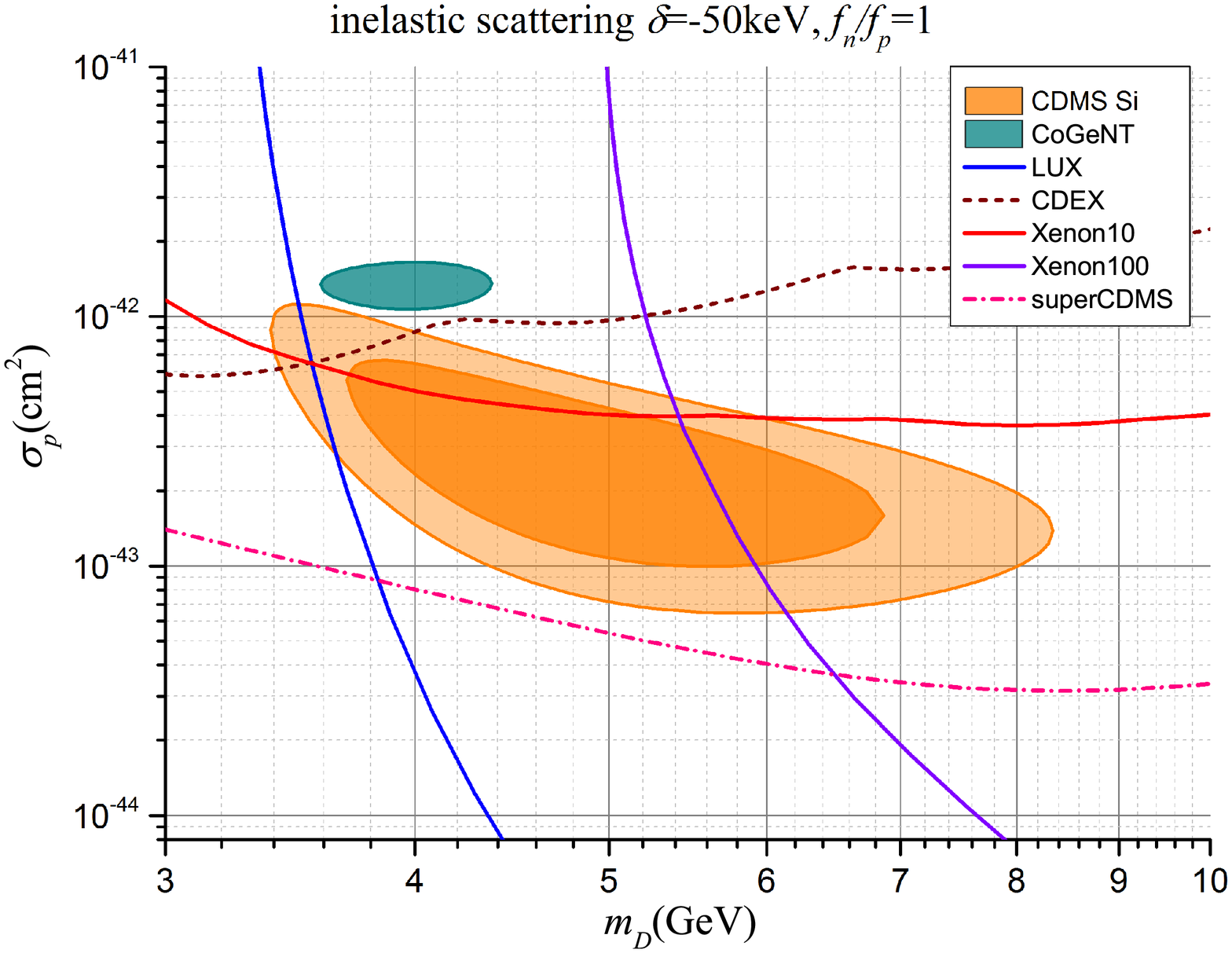}}\hspace*{-2.3cm}
\scalebox{0.4}{\includegraphics{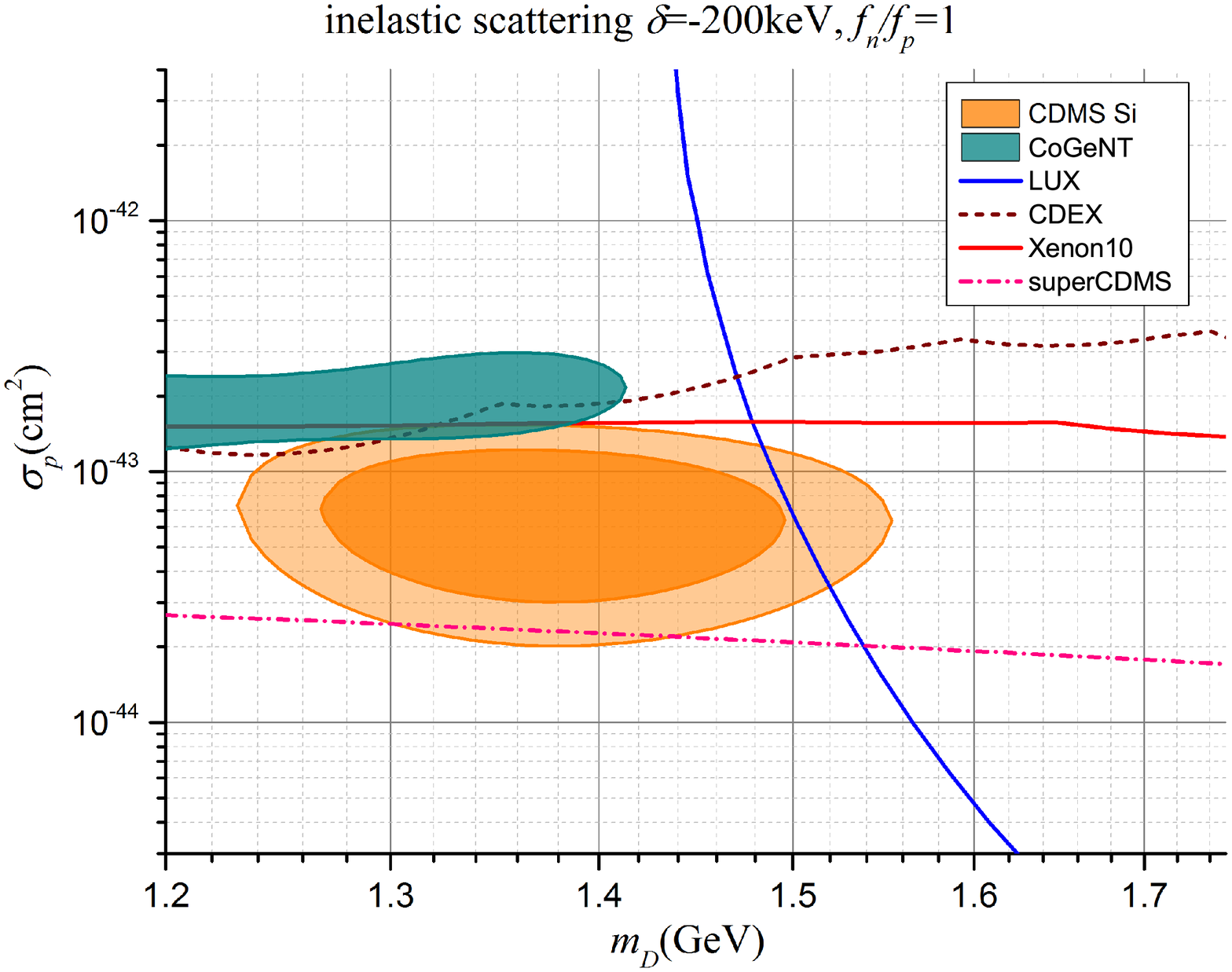}} \end{flushleft} \vspace*{-0.5cm}
\caption{Inelastic-scattering results without isospin violation. The left panel corresponds to $\delta=-50$ keV, and the right panel corresponds to $\delta=-200$ keV. For an explanation of the legend, see Fig.~1.} \end{figure}
We see that for $\delta=-50$ keV, the situation is slightly improved, whereas for $\delta=-200$ keV, the situation is much improved. The exception is CoGeNT, which is, as expected, still in tension with all null experiments; XENON10, XENON100 (lying outside the plot area to the right) and CDEX are already consistent with CDMS-Si. LUX covers almost the entire CDMS-Si contour for $\delta=-50$ keV but covers only a small portion for $\delta=-200$ keV. Only SuperCDMS still fully covers the CDMS-Si contour and is strongly in tension with the CDMS-Si result. One observation is that the signal region of CoGeNT becomes significantly larger than that for CDMS-Si at $\delta=-200$ keV. This behavior arises from the difficulty in fitting the data from the multi-events to a relatively large DM mass splitting \cite{ExcitingDM9}. The $\chi^2_{\mathrm{min}}$ of this fitting is significantly larger than that for the elastic fitting. Because $\delta=-200$ keV is already approaching the lower limit on the allowed mass difference for exothermic DM \cite{ExcitingDM7}, the results of Fig.~2 indicate that exothermic DM alone, even when an extreme $\delta$ value is used and the CoGeNT result is ignored, is still not sufficient to accommodate both the SuperCDMS and CDMS-Si results. For DAMA, note that when the inelastic scattering of DM is considered, the area of the low-mass signal region from the DAMA experiment shown in Fig.~1 reduces as the mass splitting $|\delta|$ grows. This effect can also be observed in Fig.~1 of \cite{ExcitingDM5}. In our analysis with $\delta=-50$ keV and $\delta=-200$ keV, the signal region of DAMA for low masses (masses comparable to the signal regions of CDMS-Si and CoGeNT) completely disappears, or the DAMA result shrinks to zero. For this reason, in the following, as long as we are discussing exothermic DM with $\delta=-50$ keV or $\delta=-200$ keV, we shall no longer consider the DAMA experiment. Furthermore, the inelastic-scattering DM does not fit the DAMA data well even for larger masses ($m_{\chi}>30$GeV); the $\chi^2_{\mathrm{min}}/$d.o.f is approximately 35/34 for $\delta=-200$ keV, whereas $\chi^2_{\mathrm{min}}/$d.o.f =27.8/34 for $\delta=0$.

Next, we include an isospin-violating effect. The general low-energy differential cross section is \cite{ExcitingDM3}
\begin{eqnarray}
\frac{d\sigma}{dE_R}=\frac{m_N}{2\mu^2v^2}\sigma_{\mathrm{el}}[Zf_p+(A-Z)f_n]^2F(q^2),
\end{eqnarray}
where $Z$ is the atomic number of the target nucleus; $A$ is its mass number; $f_p$ and $f_n$ are constants that represent the relative coupling strengths to protons and neutrons, respectively; and $F(q^2)$ a form factor that depends on the momentum transfer to the nucleus, $q^2 = 2m_NE_R$. $\sigma_{\mathrm{el}}$ is the elastic limit of the above cross section, which is reached when the splitting is much less than the kinetic energy of the collision. The differential scattering rate of dark matter per unit recoil energy $E_R$ is given by
\begin{eqnarray}
\frac{dR}{dE_R}=N_Tn_{\chi}\int_{v_{\mathrm{min}}}\frac{d\sigma}{dE_R}vf(v)dv,
\end{eqnarray}
where $v_{\mathrm{min}}$, which is determined using Eq. (\ref{vmin}), is the minimum velocity required to produce a recoil of energy $E_R$; $N_T$ is the number of target nuclei; $n_{\chi}$ is the local number density of the dark matter; and $f(v)$ is the distribution of DM velocities relative to the target.
With $N_Tm_N=m_{\mathrm{detector}}$ and $\rho_{\chi}=n_{\chi}m_{\chi}$, the differential recoil rate per unit detector mass can be written as
\begin{eqnarray}
\frac{dR}{dE_R}=\frac{\rho_{\chi}}{2m_{\chi}\mu^2}\sigma_{\mathrm{el}}
[Zf_p+(A-Z)f_n]^2F_A(q^2)\eta(E_R,t),
\end{eqnarray}
where $\rho_{\chi}=0.3~\mathrm{GeV}/\mathrm{c}^3$ is the local DM density. Details of the DM velocity distribution are included via the mean inverse speed $\eta(E,t)$,
\begin{eqnarray}
\eta(E_R,t)=\int_{v_{min}(E_R)}\frac{f(v)}{v}d^3v,
\end{eqnarray}
where $f(v)$ at any given time of the year is determined by the velocity of the Earth through the halo and by the distribution of DM velocities within the halo itself, here assumed to be of the form
\begin{eqnarray}
f(v)=\frac{N_0}{(\pi v_0^2)^{3/2}}e^{-v^2/v_0^2}\Theta(v_{esc}-v).
\end{eqnarray}
We have assumed a Maxwell-Boltzmann distribution for the DM halo velocities with a mean of $v_0=220\mathrm{km}/\mathrm{s}$ and a sharp cutoff (i.e., the galactic escape velocity) at $v_{esc} = 544\mathrm{km}/\mathrm{s}$. $N_0$ is chosen to normalize the probability distribution to one. Because the Earth is moving with a velocity $v_E=220\mathrm{km}/\mathrm{s}$, $\eta(E,t)$ can be written as \cite{DAMA2}
\begin{eqnarray}
\eta(E,t)=\left\{
\begin{array}{lr}
\frac{1}{v_0y}, & \mbox{for} \quad z<y,x<|y-z|,\\
\frac{1}{2N_{esc}v_0y}[\mbox{erf}(x+y)-\mbox{erf}(x-y)-\frac{4}{\sqrt{pi}}ye^{-z^2}], & \mbox{for} \quad z>y,x<|y-z|,\\
\frac{1}{2N_{esc}v_0y}[\mbox{erf}(z)-\mbox{erf}(x-y)-\frac{2}{\sqrt{pi}}(y+z-x)e^{-z^2}], & \mbox{for} \quad |y-z|<x<y+z,\\
0, & \mbox{for} \quad y+z<x,\\
\end{array}\right.
\end{eqnarray}
where
\begin{eqnarray}
x=v_{min}/v_0,\quad y=v_{E}/v_0,\quad z=v_{esc}/v_0\hspace*{2cm}
N_{esc}=\mbox{erf}(z)-\frac{2z}{\sqrt{\pi}}e^{-z^2}.
\end{eqnarray}

For the annual modulation, the count rate generally has an approximate time dependence as follows:
\begin{eqnarray}
\frac{dR}{dE_R}(E_R,t)\approx S_0(E_R)+S_m(E_R)\cos\omega (t-t_c)
\end{eqnarray}
where $t_c$ is the time of year at which $v_{obs}(t)$ is at its maximum, $S_0(E_R)$ is the average differential recoil rate over a year, and $S_m(E_R)$ is referred to as the modulation amplitude. For the standard halo model,
\begin{eqnarray}
S_m(E_R)=\frac{1}{2}\Big[\frac{dR}{dE_R}\bigg|_{v_E=v_{sun}+v_{orb}\cos\gamma}-\frac{dR}{dE_R}\bigg|_{v_E=v_{sun}-v_{orb}\cos\gamma}\Big],
\end{eqnarray}
where $v_{orb}=30\mathrm{km/s}$ and $\cos\gamma=0.51$.

Finally, to consider isospin-violating scattering from dark matter, different mass numbers will yield different differential recoil rates. The event rate is given by
\begin{eqnarray}
R=\sum_i r_i N_Tm_{A_i}\int dE_R \frac{\rho_{\chi}}{2m_{\chi}\mu^2}F_{A_i}(q^2)\eta(E_R,t),
\end{eqnarray}
where the sum is over the isotopes $A_i$ with fractional number abundances $r_i$ \cite{IVDM5}.

Using these formulae, and with a ratio of $f_n/f_p\approx-0.7$, we performed the relevant calculations, and in Fig.~3, we plot the elastic-scattering (corresponding to $\delta=0$) IVDM results of the CoGeNT, DAMA (both with and without the ion-channeling effect) and CDMS-Si signal regions, alongside the null results of XENON100, XENON10, LUX, SuperCDMS, and CDEX. Through comparison with Fig.~1, we find that the IVDM model does slightly reduce the tensions, but the null experiments LUX and  SuperCDMS are essentially still in tension with the CoGeNT, DAMA and CDMS-Si result. In particular, we recover the previously mentioned result that LUX and CDMS-Si are in tension for IVDM \cite{tension1,tension2,tension3}.

\begin{figure} \begin{center}
\scalebox{0.4}{\includegraphics{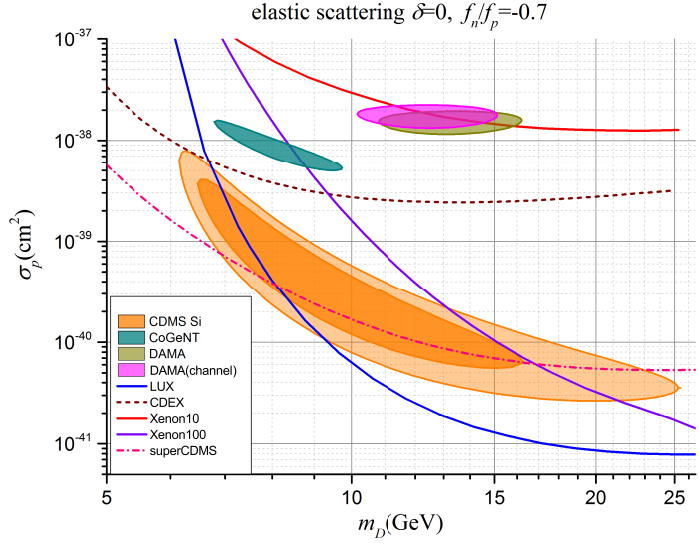}} \end{center} \vspace*{-0.5cm}
\caption{Elastic-scattering IVDM results for $f_n/f_p=-0.7$. For an explanation of the legend, see Fig.~1.} \end{figure}

We continue by considering the inelastic-scattering effects.  The underlying model for inelastic scattering is typically constructed with a vector particle--dark photon (or dark Z') mixing kinetically with an SM U(1) gauge boson and coupling to the two different DM particles, $\chi_1$ and $\chi_2$ \cite{ExcitingDM2,ExcitingDM9}; here, to ensure that the coupling of the DM particles to the dark photon is strictly off-diagonal in the mass basis, the DM particles must be Majorana states because there then exists no vector current for a single Majorana particle. In this scenario, elastic scattering between DM and nucleons can occur happen at second order (right panel of Fig.~4) and is thus suppressed, whereas inelastic scattering can occur at first order (left panel of Fig.~4) and thus plays the leading role in direct-detection experiments. If, furthermore, the kinetic energy is smaller than the mass splitting of the DM, then up-scattering on nucleons is kinetically prohibited, and we are left with the exothermic scattering of the DM.
\begin{figure} \begin{center}
\scalebox{0.6}{\includegraphics{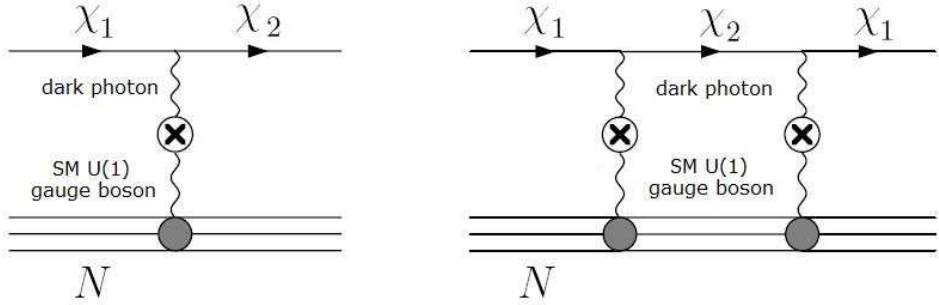}} \end{center} \vspace*{-0.5cm}
\caption{First- and second-order Born amplitudes for DM-nucleus scattering \cite{ExcitingDM2}.} \end{figure}
To further account for the large isospin-violating effect, the conventional Higgs portal scheme of a scalar field mixing with the SM Higgs to communicate between the SM and DM sectors causes no significant isospin violation\cite{Z'IVDM1} because only a very small percentage of the nucleon constituents are related to the current quark mass and thus connected to the Higgs field. We then must exploit vector instead of scalar particles to connect the dark world with SM particles \footnote{Indeed, we have investigated the possibility that instead of treating the extra vector boson as a messenger that connects the DM world with SM particles, we may treat it merely as a single DM candidate \cite{Z'SingleDM}. The result reveals that DM of this type must have a mass larger than the weak W boson mass and therefore is unrelated to the present GeV DM, which is a possibility that supports the present choice of a messenger role for the vector boson in our low-energy-region search for DM.}. For such a model with a single messenger, the isospin-violating effect depends on the choice of SM U(1) with which the new vector boson mixes. For example, if, as usual, we take U(1) to be the SM hypercharge $\mathrm{U(1)_Y}$ \cite{ExcitingDM9}, because the proton and neutron have the same hypercharge, we then expect the plot for the left diagram of Fig.~4 to be the same for both neutrons and protons, leading to $f_n=f_p$, i.e., there is no isospin violation. If, instead, we consider that in the low-energy region, the $Z$-boson component of $\mathrm{U(1)_Y}$ decouples, then effectively, only the electromagnetic part will contribute, and we can then take U(1) to be the SM electromagnetic $\mathrm{U(1)_{em}}$ \cite{ExcitingDM2,ExcitingDM7}; because the neutron is neutral and the proton is charged, we then expect the same plot to be zero for neutrons, resulting in $f_n=0$ and $f_p\neq 0$, i.e., we have isospin violation. In Fig.~5, we plot the $f_n=0$ IVDM exothermic DM result for the CoGeNT and CDMS-Si signal regions along with the null results of XENON10, LUX, SuperCDMS, and CDEX (XENON100 lies to the right, outside the plot area).
\begin{figure} \begin{center}
\scalebox{0.4}{\includegraphics{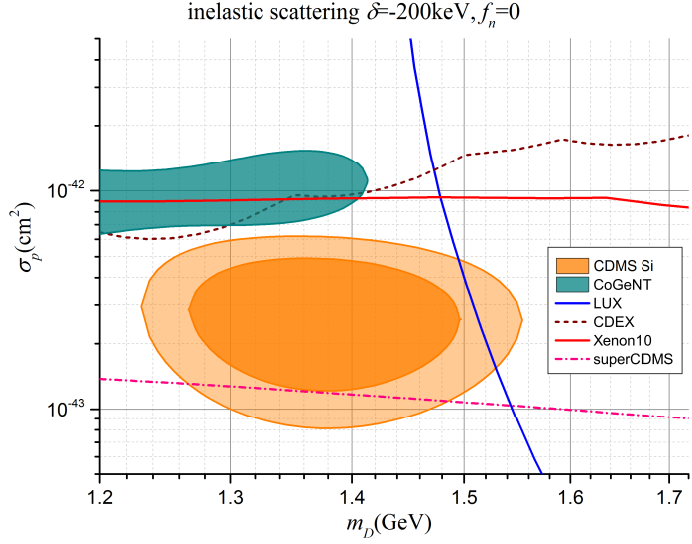}} \end{center} \vspace*{-0.5cm}
\caption{Inelastic IVDM scattering result for $f_n=0$ and $\delta=-200$ keV. For an explanation of the legend, see Fig.~1.} \end{figure}

Comparison of the plots of Fig.~5 and the right panel of Fig.~2 reveals only a very few changes. In particular, the SuperCDMS result is only marginally in tension with the CDMS-Si result. This is because the maximum suppression values of $f_n/f_p$ are $-0.785$ (for Ge), $-0.697$ (for Xe), and $-0.992$ (for Si), and a detailed computation shows that if we take $f_n/f_p=-0.7$, then the energy spectra of Ge and Xe relative to Si are suppressed by approximately 90$\%$ and 95$\%$, respectively; if we set $f_n=0$, then the suppression of Ge and Xe relative to the Si energy spectra is reduced by 20$\%$. Hence, $f_n=0$ offers an insufficient isospin-violating effect, and we must increase its strength by setting $f_n/f_p=-0.7$. In the literature, the first discussion of vector boson exchange leading to $f_n/f_p=-0.7$ was presented in Ref.~\cite{Z'IVDM2}, and in that discussion, the key roles were played by three factors: the conventional kinetic mixing and the mass mixing between SM $U(1)$ and the dark photon or Z' as well as the coupling of the dark photon to SM quarks.  Although the original model presented in Ref.~\cite{Z'IVDM2} does not include the inelastic-scattering effect, we can modify the model by adding a Majorana mass term to the DM fields, which will yield exactly an off-diagonal dark-photon coupling to the DM fermions, and this improvement does not change the value of $f_n/f_p$. To be more explicit, we write a Lagrangian for our proposed schematic model as follows:
\begin{eqnarray}
&&\hspace*{-1cm}\mathcal{L}=\mathcal{L}_{\mathrm{SM}}-\frac{1}{4}X^{\mu\nu}X_{\mu\nu}
+\frac{1}{2}m_X^2X_\mu{X}^\mu-m{_\chi}\bar{\chi}\chi-\frac{1}{2}\sin\epsilon\;{B}_{\mu\nu}X^{\mu\nu}
+\delta m^2Z_\mu{X}^\mu~~~~\label{OurModel}\\
&&+\bar{\chi}(i\slashed{\partial}-f_{\chi}^V\slashed{X})\chi-{\displaystyle\sum_f}f_f^V\bar{f}\slashed{X}f
-\frac{\delta}{2}(\bar{\chi}^c\chi+\bar{\chi}\chi^c),\nonumber
\end{eqnarray}
where the extra $U(1)_{\mathrm{X}}$ is assumed to be broken and the corresponding vector boson mass is $m_{Z'}$. We denote the fields in the interaction basis by $(B,W^3,X)$
 and in the mass-eigenstate basis by $(A,Z,Z')$, and we define $Z\equiv{c}_W{W}^3-s_W{B}$, where $s_W$($c_W$) is the sine (cosine) of the Weinberg angle. $\chi$ is the fermionic DM field with Dirac mass $M$ and Majorana mass $\delta\ll M$.

For this Lagrangian, the discussions of Ref.~\cite{Z'IVDM2} demonstrate that there exist suitable parameter spaces $(\epsilon,\delta m^2,f_f^V)$ to account for $f_n/f_p=-0.7$, as described in greater detail below.
\begin{itemize}
\item For the dark Z' scenario, in which the SM fields are uncharged under the extra $U(1)_{\mathrm{X}}$ group and, thus, $f_f^V=0$, Fig.~2 of Ref.\cite{Z'IVDM2}
shows that the ratio $f_n/f_p\sim 0.7$ with $m_{Z'}=4$ GeV can be achieved by adjusting the remaining two parameters $\epsilon$ and $\delta m^2$ appropriately.
The figure shows that for $\epsilon\approx\delta m^2/m_Z^2$ and $\epsilon\ll 1$, we have $f_n/f_p\approx 1/3s_W\approx-0.7$.
\item For the baryonic Z' scenario, the SM is charged under the $U(1)_{\mathrm{X}}$ group, whereas the leptons are uncharged under  $U(1)_{\mathrm{X}}$ and  $U(1)_{\mathrm{X}}\equiv U(1)_{\mathrm{B}}$. In this case, $f_u^V=f_d^V\equiv f_q^V$. Now, there are three parameters, $(\epsilon,\delta m^2,f_f^V)$. Figure.~3 of Ref.\cite{Z'IVDM2} shows that the ratio $f_n/f_p\sim 0.7$ can be achieved by adjusting two of the three parameters; the left panel illustrates the variation of $\epsilon$ and $f_q^V$ with $\delta m^2=0$, and the right panel illustrates the variation of $\epsilon$ and $\delta m^2$ with $f_q^V\approx 10^{-5}$. The figure shows that to obtain $f_n/f_p\approx -0.7$, $f_q^V$ must be more than an order of magnitude smaller than $\epsilon$. Suppose that $\epsilon$ in Ref.~\cite{Z'IVDM2} is constrained to be on the order of $10^{-2}$ or smaller, such that $f_q^V\le 10^{-3}$. The requisite smallness of $f_q^V$ may be achieved by coupling $Z'$ only to the second- and third-generation quarks, and this relaxes the restriction that the additional $U(1)_{\mathrm{X}}$ must be baryonic, thereby allowing for couplings to leptons to facilitate the construction of an anomaly-free model.
\end{itemize}
By contrast, diagonalizing the DM mass matrix leads to mass eigenstates $\chi_{1,2}$ of masses $M_{1,2}=m_\chi\mp\delta$ and an off-diagonal gauge interaction, which leads to the DM scattering picture previously considered in Fig.~4.
\begin{eqnarray}
\bar{\chi}(i\slashed{\partial}-f_{\chi}^V\slashed{X})\chi=
\bar{\chi}_1i\slashed{\partial}\chi_1+\bar{\chi}_2i\slashed{\partial}\chi_2-f_{\chi}^V(\bar{\chi}_1\slashed{X}\chi_2+\bar{\chi}_2\slashed{X}\chi_1).
\end{eqnarray}

Ref.~\cite{Z'IVDM3} presents similar discussions with two additional important extensions: first, noting the possibility of applying the model to inelastic scattering, and second, proving that a combination with the conventional Higgs mediator will help to achieve the desired isospin violation. These extensions are also investigated in Ref.~\cite{Z'IVDM1}, and the combination of the dark photon and the conventional Higgs mediator is further generalized to the combination of the dark photon and another new light vector gauge boson. Using our schematic model (\ref{OurModel}), especially the parameter range represented in Fig.~2 and Fig.~3 of Ref.\cite{Z'IVDM2}, in addition to these other possible underlying exothermic IVDM models that give rise to an expected value of $f_n/f_p=-0.7$, we plot the IVDM exothermic DM result for the CoGeNT and CDMS-Si signal regions along with the null results of XENON100, XENON10, LUX, SuperCDMS, and CDEX (see Fig.~6). The left plot corresponds to $\delta=-50$ keV, and the right plot corresponds to $\delta=-200$ keV.
\begin{figure} \begin{flushleft}
\hspace*{-1.3cm}\scalebox{0.4}{\includegraphics{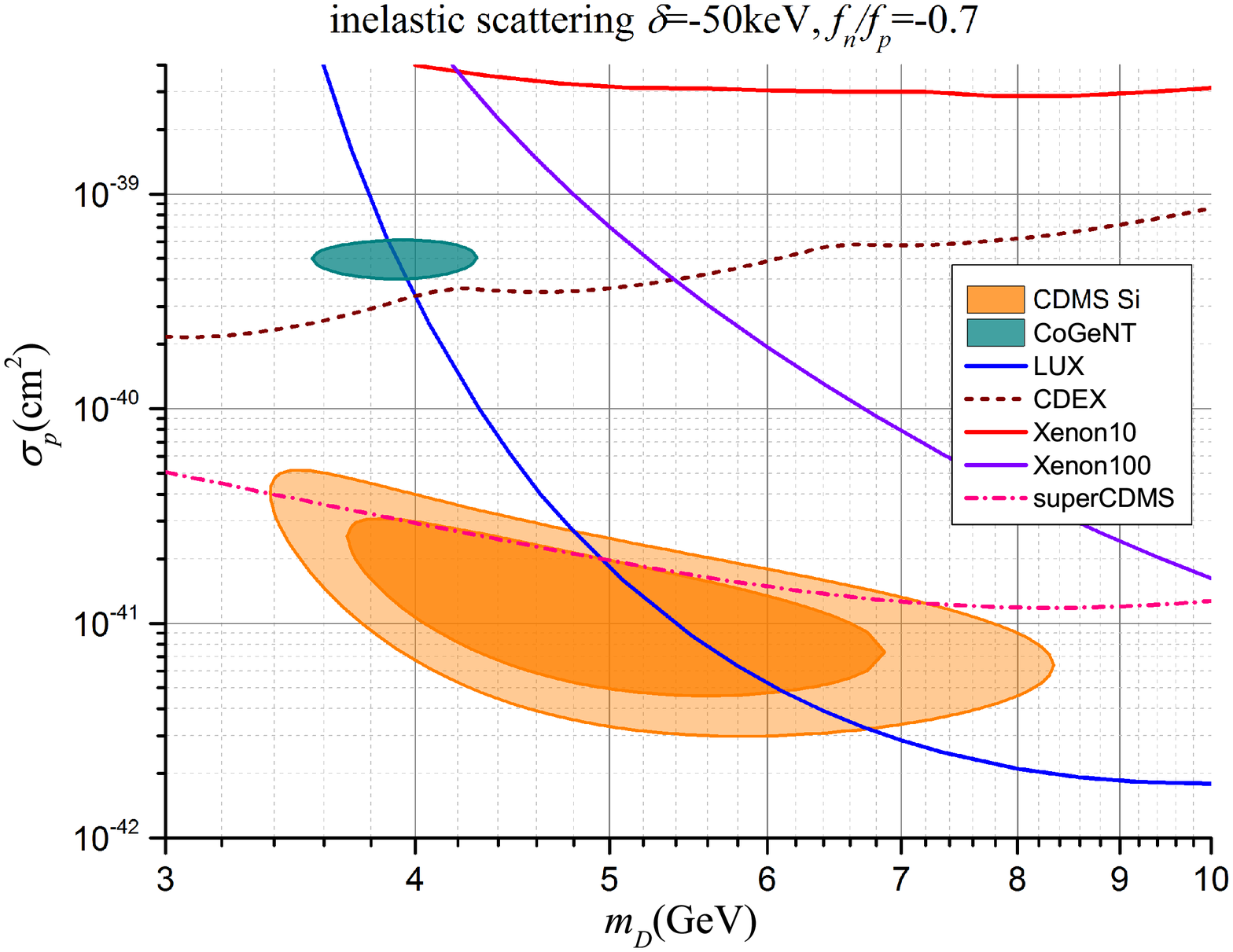}}\hspace*{-2.3cm}
\scalebox{0.4}{\includegraphics{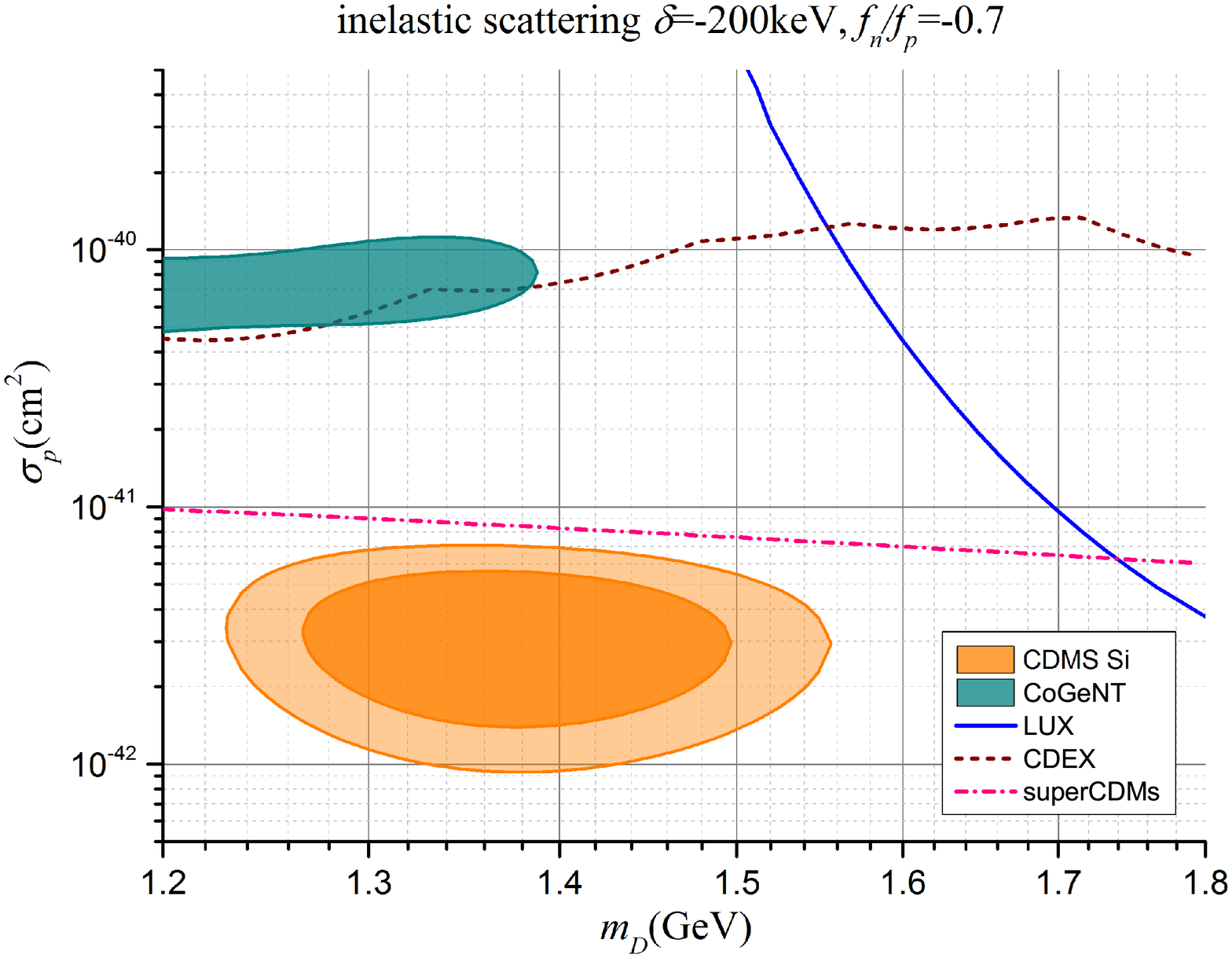}} \end{flushleft} \vspace*{-0.5cm}
\caption{Inelastic IVDM scattering results for $\delta=-50$ keV (left panel) and $\delta=-200$ keV (right panel). For an explanation of the legend, see Fig.~1.} \end{figure}
Apart from the strong tension remaining between CoGeNT and the null experiments, we see that even for $\delta=-50$ keV, CDMS-Si is already consistent with most of the null experiments, although LUX cuts through half of the contour. For $\delta=-200$ keV, the tensions between CDMS-Si and the null experiments are over-relaxed. Therefore, with the assistance of isospin-violating effects from the dark photon or Z', we can readily make CDMS-Si consistent with all current null experiments, even without invoking the extreme case of exothermic DM with $\delta=-200$ keV. This leaves open a region in the parameter space for exothermic DM to fit other current and future DM detection experiments.

It should be noted that there are several other possible methods of reconciling the tensions among various direct-detection experiments. The first is to interpret the possible signals appearing in DAMA, CoGeNT, and CDMS-Si not as DM signals but as some atmospherically produced neutral particle with a relatively large magnetic dipole moment \cite{MdipoleMoment}, as such particles can mimic DM signals. A very definite flux could explain the signals observed in DAMA/LIBRA, CDMS-Si, and CoGeNT that are consistent with the bounds from XENON100 and CDMS-II. In this scenario, the key is that the recoil energy of the assumed particle must lie some specific energy range that is above the thresholds of DAMA/LIBRA, CDMS-Si, and CoGeNT but below those of XENON100 and CDMS-II. If we further consider the latest results of SuperCDMS and CDEX, then this recoil energy must lie above the thresholds of these two experiments and therefore is expected to produce signals in these detectors. This has is not occurred, hence implying that this alternative interpretation is not favored by the latest SuperCDMS and CDEX null results.

The second possibility is to invoke composite DM, wherein stable particles of charge 2 bind with primordial helium to form O-helium "atoms" (OHe), representing a specific warmer-than-cold nuclear-interacting form of dark matter \cite{OHe}. Because it slows down in terrestrial matter, OHe is elusive in direct methods of underground DM detection such as those used in the CDMS experiment, but its reactions with nuclei can lead to annual variations in the energy released in the energy range of $2-6$~keV such as those observed in the DAMA/NaI and DAMA/LIBRA experiments. However, this class of solution cannot explain the unmodulated signals in experiments such as CoGeNT and CDMS-Si and therefore is not favored by these experiments.

Finally, for completeness, we will list for each experiment some of the details of the computations used to obtain all the above plots (except Figs.~4):
\begin{enumerate}
\item\underline{\bf CDMS-Si}:~~We used the acceptance from \cite{CDMSSi} and a total exposure of 140.2 kg-days, assuming zero background. We considered an energy interval of [7,100] keV and binned the data in 2 keV intervals as in \cite{tension3}. The three candidate events appeared in the first three bins.
To find the best-fit regions, we obtained the extended log-likelihood function and simply plotted constant values of the likelihood that it would correspond to the 68$\%$ and 95$\%$ CL regions under the assumption that the likelihood distribution is Gaussian.
\item\underline{\bf CoGeNT}:~~We used the data and flat background shown in Fig.~23 of \cite{Cogent3}, which has been corrected for efficiency (i.e., bin counts have been scaled to reflect the numbers of events expected based on those observed and the deduced efficiency). We performed a $\chi^2$ scan over a cross section using the DM mass and background from the data of Ref.~\cite{Cogent3}. The curves for the region of interest correspond to the 90$\%$ C.L. regions. The energy resolution below 10 keV was taken to be that reported by CoGeNT, namely, $\sigma^2=\sigma_n^2+ 2.35^2E\eta F$, where $\sigma_n=69.4$ eV is the intrinsic electronic noise, $E$ is the energy in eV, $\eta=2.96$ eV is the average energy required to create an electron-hole pair in Ge at approximately 80 K, and $F=0.29$ is the Fano factor. The number of expected events in a given range was taken to be \cite{tension1}
\begin{eqnarray}
N_{[E_1,E_2]}=Ex.\int_0^{\infty}\frac{dR}{dE_R}\mbox{res}(E_1,E_2,E_R)dE_R+b_{[E_1,E_2]},
\end{eqnarray}
where $b$ is the flat, floating background and $2\mbox{res}(E_1, E_2; E_R)=\mbox{erf}((E_1-E_R)/ (\sqrt{2}\sigma))-\mbox{erf}((E_2-E_R)/(\sqrt{2}\sigma))$.
\item\underline{\bf DAMA}:~~The average amplitude over the energy interval $[E_1,E_2]$ is
\begin{eqnarray}
S_{m,[E_1,E_2]}=\frac{1}{E_2-E_1}\sum_{T=Na,I}c_T\int_{E_1/Q_T}^{E_2/Q_T}S(E_R)dE_R,
\end{eqnarray}
where $c_T$ is the mass fraction of the target and $Q_T$ is the quenching factor for the target, which we take to be $Q_{Na}=0.3$ and $Q_I=0.09$. To account for the ion-channeling effect, we take the channeling fraction to be
\begin{eqnarray}
f_{Na}=10^{-\sqrt{E/(6.9\mathrm{keV})}},\qquad f_{I}=10^{-\sqrt{E/(11.5\mathrm{keV})}},
\end{eqnarray}
as in Ref.~\cite{DAMA6}. The measured energy will be normally distributed with a standard deviation of
\begin{eqnarray}
\sigma(E)=(0.448\mathrm{keV})\sqrt{E/\mathrm{keV}}+0.0091E.
\end{eqnarray}

We used the data presented in Fig.~6 of \cite{DAMA3}. We calculated $\chi^2$ using all 36 bins corresponding to energies from 2 keV to 20 keV. The 95$\%$ C.L. contours of the region of interest satisfy $\chi^2=\chi^2_{min}+\mathrm{CDF}^{-1}(\mathrm{ChiSq[2],C.L.}).$

\item\underline{\bf XENON10}:~~We simply adopted the collaboration's parameterization from Fig.~1 of \cite{XENON10}, assuming a sharp cutoff to zero at a nuclear recoil energy of 1.4 keV. The signal region is from 5 to 35 electrons, corresponding to nuclear recoils of 1.4 keV to 10 keV. A limit of 90$\%$ C.L. was obtained using the $p_{\mbox{max}}$ method \cite{Pmax} and the 23 highlighted S2 event signals from Fig.~2 of \cite{XENON10}.
\item\underline{\bf XENON100}:~~We used the mean $\nu(E)$ characterized in \cite{XENON1004}. For the scintillation efficiency $\mathcal{L}_{eff}$, we used the efficiency used in XENON100's 225-live-day analysis \cite{XENON1003} obtained from Fig.~1 of \cite{XENON1001}, which included a linear extrapolation to 0 for $E$ below 3 keV. The response of the detector was modeled as a Gaussian distribution with a mean of $n$ and a variance of $\sqrt{n}\sigma_{PMT}$, where $\sigma_{PMT}=0.5 \mbox{PE}$ \cite{XENON1004}. The Gaussian smearing also included a photoelectron-dependent acceptance, which we parameterized based on Fig.~1 of \cite{XENON1003}. To obtain the total rate, we summed the differential rate over the signal region, which corresponds to $S1\in(3, 30) \mbox{PE}$ for the analysis presented in \cite{XENON1003}, and used a total exposure of 225$\times$34 kg-days \cite{XENON1003}.
We then used Poisson statistics to obtain a 90$\%$ C.L. upper limit, where two events were observed, as shown in Fig.~2 of \cite{XENON1003}.
\item\underline{\bf LUX}:~~The experimental design of LUX is quite similar to that of XENON100 \cite{ExcitingDM9}. Both experiments use a combination of scintillation (S1) and ionization signals (S2) to effectively reject background. Following \cite{XENON1004}, we computed the number of signal events as follows:
\begin{eqnarray}
N_{DM}=\int_{S1_{lower}}^{S1_{upper}}dS1\sum_{n=1}^{\infty}
\mbox{Gauss}(S1|n,\sqrt{n}\sigma_{PMT})\int_0^{\infty}dE_R\epsilon(E_R)
\mbox{Poisson}(n|\nu(E_R))\frac{dR}{dE_R}\times\mbox{Ex.},
\end{eqnarray}
where Ex. denotes the experimental exposure, $\epsilon(E_R)$ is the S1 efficiency, and $\sigma_{PMT}=0.37~\mbox{PE}$ accounts for the PMT resolution. For the LUX analysis, $S1_{lower}=2$ and $S1_{upper}=30$. The expected number of photoelectrons $\nu(E_R)$ is
\begin{eqnarray}
\nu(E_R)=E_R\times \mathcal{L}_{eff}(E_R)\times\frac{S_{n}}{S_{e}}\times L_y,
\end{eqnarray}
where $\mathcal{L}_{eff}$ is the energy-dependent scintillation efficiency of liquid xenon, $L_y$ is the light yield, and $S_n$ and $S_e$ are the nuclear- and electron-recoil quenching factors, respectively, that arise from the applied electric field. We used the energy-dependent absolute light yield, $\mathcal{L}_{eff}(E_R) \frac{S_{n}}{S_{e}}L_y$, from slide 25 of \cite{LUX1}, with a hard cutoff below 3 keV. Finally, for the DM detection efficiency, we used the efficiency calculated after threshold cuts from Fig.~9 of \cite{LUX}. We computed 90$\%$ CL limits using Poisson statistics with no events detected.
\item\underline{\bf SuperCDMS}:~~For the efficiency reported in Fig.~1 of \cite{SuperCDMS}, we used the 577-kg-day data from \cite{SuperCDMS}. To obtain the 90$\%$ C.L. limits, we used Poisson statistics with 11 candidate events detected, which are listed in Table 1 of \cite{SuperCDMS}, and zero background was assumed.
\item\underline{\bf CDEX}:~~We assumed perfect efficiency and used the 53.9-kg-day data from the residual spectrum presented in Fig.~3(b) of \cite{CDEX2}. A flat background was assumed, as given by the minimum $\chi^2$ method. The quenching factor of a recoiling Ge nucleus was obtained from the TRIM program as in \cite{CDEX3}. To obtain the 90$\%$ C.L. limits, the binned Poisson method \cite{DAMA2} with bins of $0.1$~keVee was used.
\end{enumerate}

To summarize, we find that exothermic DM alone is not sufficient to fully resolve the tensions between CDMS-Si and the null experiments. However, if some underlying interaction allows isospin-violating effects to be incorporated into exothermic DM models, then, with the aid of the strongest setting $f_n/f_p=-0.7$, exothermic IVDM can make the CDMS-Si result consistent with the results of all the latest null experiments, except the CoGeNT experiment. Meanwhile,
for exothermic IVDM to exist, at least a new vector gauge boson (dark photon or dark Z') that connects SM quarks with Majorana-type DM particles is required.
 \section*{Acknowledgments}
 This work was supported by the National Basic Research Program of China (973 Program) under Grant No. 2010CB833000, the National Science Foundation of China (NSFC) under Grant No. 11475092, the Specialized Research Fund for the Doctoral Program of High Education of China No. 20110002110010, and the Tsinghua University Initiative Scientific Research Program No. 20121088494.

\end{document}